\documentclass[prl,twocolumn,showpacs]{revtex4}
\usepackage{graphics}
\usepackage{color}
\usepackage{amsmath,amssymb,amsthm}
\usepackage{times}
\definecolor{wine}{rgb}{0.5,0.1,0.1}
\begin{document}

\title{lattice Boltzmann simulation of electrowetting-on-dielectric in a rough-wall channel}

\author{Huabing Li$^{1,2}$, Jian Li$^{2}$, YanYan Chen$^{1,3}$ and
  Haiping Fang$^{1}$\footnote{To whom correspondence should be
  addressed. Email address: fanghaiping@sinap.ac.cn}
}

\affiliation{
  $^1$Shanghai Institute of Applied Physics, Chinese Academy of
  Sciences, P.O. Box 800-204, Shanghai 201800, China\\
  $^2$Department of information material science and engineering,
  Guilin University of Electronic Technology, Guilin 541004, China\\
  $^3$College of Mathematics, Physics and Information Engineering,
Zhejiang Normal University, Jinhua 321004, China }

\begin{abstract}

A lattice Boltzmann model was proposed  to simulate
electrowetting-on-dielectric (EWOD). The insulative vapor and the
electrolyte liquid droplet were simulated by the lattice Boltzmann
method respectively, and the linear property between cosine of
contact angle and the electric field force confirms the
reliability of this model. In the simulation of electrolyte
flowing in a rough-wall channel under an external electric field,
we found that a narrow channel is more sensitive than a broad
channel and the flux decreases monotonously as the electric field
increase, but may suddenly increase if the electric field is
strong enough.

\end{abstract}


\maketitle

  With the development and progress of nano-science and nano-technology, how
to manipulate tiny amounts of liquids on surfaces becomes great
challenge to nano-engineering. Gradients of both temperature and
concentration of surfactants across droplets may produce forces to
propel droplets~\cite{Mugele05}, but compared to these methods,
chemical and topographical structuring of surfaces are much finer
for controlling the equilibrium morphology of a droplet. To
further actively control the wettability of a droplet, we may coat
the droplet with self-assembled monolayers~\cite{Mugele05}, but
considering the switch speed, range of contact angle variation,
and long term stability, electrowetting (EW) may be the best
method to control a droplet on a surface.

  In electrowetting, electrowetting on dielectric (EWOD)~\cite{Berge93} and
electrowetting on insulator coated electrodes
(EICE)~\cite{Quilliet01} must be the two widely used methods to
control a droplet, because of the several advantages: without any
electrochemical reactions, clean surfaces ensure reversibility,
and highly nonwetting surfaces may be used to enhance
electrowetting effect~\cite{Squires05}. Recently, the experimental
research on electrowetting has made great progress: electrowetting
has been promoted to aplication of
lab-on-a-chip~\cite{Srinivasan04}; through an external voltage,
adjusting hydrophobic nature of an oil droplet containing
dissolved dye, an electrowetting-based reflective display has been
invented~\cite{Hayes03}; changing the contact angle of a sessile
droplet via electrowetting, shape of the droplet, as well as the
focal length of this droplet can be changed, and a flexible liquid
lenses may be made \cite{Berge00}. Theoretically, the contact
angle should decrease rapidly as the external voltage increases,
but in experiment, the contact angle has been found to be
saturated when the external voltage is high enough~\cite{Berge93}.
Because many factors may affect the contact angle of a droplet in
experiment, the saturation mechanism of this angle keeps
unknown~\cite{Mugele05}, despite several various mechanisms have
been proposed~\cite{Mugele05}. In order to study the variation of
contact angle on electrowetting as well as the electric effect on
microchannel and microfluidics, it is desirable to propose a
numeric method to study electrowetting of tiny electrolyte droplet
on an insulative layer.

   Since the lattice Boltzmann method
(LBM)~\cite{Qian92,Schen91} has been proposed, it has succeeded in
simulating complex fluid flows including multi-phase flow
\cite{Shan93} particle suspension flow~\cite{Ladd94,Li04}, binary
mixture~\cite{Shan93,Gunst91} and blood
flow~\cite{Buick02,Fang02,Migliorini02}. Because of its
intrinsically parallel and simpleness, it has been recognized as
an alternate method for computational fluid dynamics
\cite{Chen98}. Through introducing intermolecular forces, the
lattice Boltzmann method may be used to study microfluidics
\cite{Baoming03}, microchannel and nanochannel flows
\cite{Sbragaglia06}.

In this Letter, base on the multiple phases lattice Boltzmann
 model~\cite{Shan93}, considering the electric force exerted on the
 electrolyte near an insulative layer, a lattice Boltzmann model was
 proposed to investigate two-dimensional electrolyte electrowetting on a smooth
 insulative layer and the effect of EWOD on the electrolyte flow in a rough-wall channel.

In two dimensions, the system of the D2Q9 lattice Boltzmann
model~\cite{Qian92} is described by a single particle distribution
function $f_\alpha({\bf x},t)$, which can be thought as the number
of fluid particles at site ${\bf x}$, time $t$, and possessing one
of the nine velocities ${\bf e}_\alpha$ with ${\bf e}_0 = (0,0)$,
${\bf e}_\alpha = (\cos \pi (\alpha-1)/2$, $\sin \pi
(\alpha-1)/2),\alpha=1, 2, 3, 4 $, and ${\bf e}_\alpha =
\sqrt{2}(\cos \pi (2\alpha-1)/4, \sin \pi (2\alpha-1)/4)$, for
$\alpha=5, 6, 7, 8$.
 The distribution function evolves according to
a Boltzmann equation that is discrete in both space and time,

\begin{equation}
  f_\alpha({\bf x}+{\bf e}_\alpha,t+1)-f_\alpha({\bf x},t) = -\frac 1\tau(f_\alpha-f_\alpha^{eq}).
  \label{eq}
\end{equation}
The density $\rho$ and macroscopic velocity ${\bf u}$ are defined
as
\begin{equation}
  \label{rho}
  \rho = \sum_\alpha f_\alpha, \ \ \ \rho {\bf u} = \sum_\alpha f_\alpha{\bf e}_\alpha.
\end{equation}
In order to reproduce the Navier-Stokes equation, the equilibrium
distribution function $f_\alpha^{eq}$ may have the following
form~\cite{Qian92}

\begin{eqnarray}
f_\alpha^{eq} =w_\alpha\rho[1+3{\bf e}_\alpha\cdot{\bf u}
+ {\frac{9}{2}}({\bf e}%
_\alpha\cdot{\bf u})^2-{\frac{3}{2}}u^2],  \label{eqf}
\end{eqnarray}
where  $w_0 = 4/9$, $w_\alpha = 1/9$, for $\alpha$ = 1, 2, 3, 4,
and $w_\alpha = 1/36$, for $\alpha$ = 5, 6, 7, 8.

Considering the intermolecular interaction, through an interaction
potential among particles in the LBM, Shan and Chen improved the
LBM to simulate multiple phases fluid~\cite{Shan93}:
\begin{equation}
V(\textbf{X},\textbf{X}')=G_{\alpha}(\textbf{X},\textbf{X}')\varphi(\textbf{x})\varphi(\textbf{x}'),
\end{equation}
where $G_{\alpha}(\textbf{X},\textbf{X}')$ is a Green function and
$G_{\alpha}(\textbf{X},\textbf{X}') = \cal{G}$ for
$|\textbf{X}-\textbf{X}'| = \sqrt{2}$,
$G_{\alpha}(\textbf{X},\textbf{X}') = 2\cal{G}$ for
$|\textbf{X}-\textbf{X}'| = 1$ and
$G_{\alpha}(\textbf{X},\textbf{X}') = 0$ otherwise. Here we take
$\varphi(\textbf{x}) = \rho_0(1-exp(-\rho/\rho_0))$ with $\rho_0$
of a constant.
  The resulting force is
\begin{equation}
    \textbf{F}_{\sigma}(\textbf{x}) = -\varphi(\textbf{x})\sum_\alpha
    G_{\alpha}\varphi(\textbf{x}+\textbf{e}_\alpha)\textbf{e}_\alpha.
\end{equation}
If the node at $(\textbf{x}+\textbf{e}_\alpha)$ is  a solid node,
similar to a fluid node, the according resulting force can be
written as~\cite{Martys96}
\begin{equation}
    \textbf{F}_{w}(\textbf{x}) = -\varphi(\textbf{x})\sum_\alpha
    W_{\alpha} S(\textbf{x}+\textbf{e}_\alpha)\textbf{e}_\alpha,
\end{equation}
where $W_{\alpha} = w, $ for $|\textbf{e}_\alpha| = \sqrt{2}$,
$W_{\alpha} = 2w$ for $|\textbf{e}_\alpha| = 1$ and $W_{\alpha} =
0$ for $|\textbf{e}_\alpha| = 0$. If the node at
$(\textbf{x}+\textbf{e}_\alpha)$ is  a solid node, $S = 1$, else
$S = 0$.

    For an EWOD system, which consists of a conductive droplet and a metallic electrode
    insulated with an insulative layer, the double conductive layer
    has a fixed capacitance per unit area, $c =
    \varepsilon_0\varepsilon_d/d$. $d$ and $\varepsilon_d$ are the thickness and dielectric constant of
    the insulative layer. $\varepsilon_0$ is the vacuum permittivity. The free energy of the system
    reads
    \begin{equation}
    F^e =
    A_{lv}\sigma_{lv}+A_{sv}\sigma_{sv}+A_{sl}(\sigma_{sl}-\frac{\varepsilon_{0}\varepsilon_dU^{2}}{2d})-\triangle
    p V,
    \end{equation}
    where $A$ is the area of the interface and $\sigma$ is the
    interfacial tension. The subscript $lv$, $sv$, and $sl$
    indicate the liquid-vapour, solid-vapour, and solid-liquid
    interfaces respectively. $U$ is the potential difference
    between the conductive droplet and the metallic electrode.
    $\triangle p$ describe the pressure drop across the liquid-vapour interface
    and $V$ is the volume of the droplet. Through minimizing the
    free energy $F^e$, we obtain the basic equation of
    EWOD~\cite{Mugele05}
    \begin{equation}
    \cos(\theta)-\cos(\theta_0) = \frac{\varepsilon_0\varepsilon_d
    U^2}{2d\sigma_{lv}}, \label{be}
    \end{equation}
    where $\theta$ is the contact angle of the droplet and $\theta_0$
    is that without external electric feild.
    The electric force exerted on the bottom layer of the conductive
    droplet per unit area is $\textbf{f} = f \textbf{n}=
    \frac{\varepsilon_0{\varepsilon_d}U^2}{2d}\textbf{n}$ and $\textbf{n}$
    is the unit normal vector of the layer pointing to the metallic electrode. Thus equation
    (\ref{be}) is simplified as
\begin{equation}
\cos(\theta)-\cos(\theta_0) = \frac{f}{\sigma_{lv}}. \label{be1}
\end{equation}
Based on equation (\ref{be1}), through a force exerted on the
single layer of the liquid near the insulative layer, we may
simulate an electrolyte droplet electrowetting on an insulative
layer by the LBM.

\begin{figure}[htb]
   \centering
   {\scalebox{0.7}[0.7]{\includegraphics*[220,430][400,540]
        {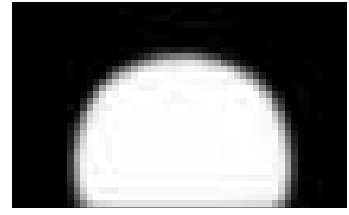}}}
  \caption{Shape of a droplet without external electric field.}
  \label{droplet}
\end{figure}

\begin{figure}
\centering
   {\scalebox{0.35}[0.35]{\includegraphics*[60,240][530,600]
        {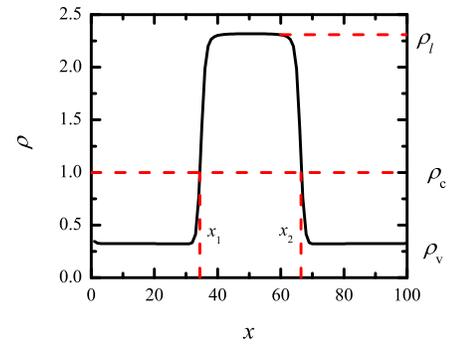}}}
  \caption{The density of the single layer of fluid near the substrate.
  $\rho_l$ and $\rho_v$ are the density of the conductive liquid and
  insulative vapor respectively. $\rho_c$ is a critical density to
  distinguish whether the fluid is conductive ($\rho \geq
  \rho_c$) or insulative ($\rho < \rho_c$). The fluid in the
  range of $x_{1} \leq x \leq x_{2}$ is regarded as conductive.
  Here we choice $\rho_c = 1.0.$}\label{density}
  \label{angle}
\end{figure}

Fig. \ref{droplet} is a droplet adopted on a smooth substrate
simulated by the LBM. The single layer of fluid near the substrate
is shown in Fig. \ref{density}. The fluids in the center with a
high density $\rho_l$ consist the liquid, the other fluids in the
two sides with a low density $\rho_v$ belong to the vapor. There
is a transition part connect the liquid and the vapor. Because the
liquid is conductive but the vapor is insulative, it is necessary
to introduce a critical density $\rho_c$ ($\rho_v < \rho_c
<\rho_l$) to distinguish the fluid from conductor (with $\rho \geq
\rho_c$) and insulator (with $\rho <\rho_c$). In Fig.
\ref{density}, the fluids in the range between $x_1$ and $x_2$ in
the single layer near the substrate are conductive. Fluids outside
this range are regarded as insulative vapor.

Under the influence of such forces, the momentum at each fluid
node is changed to
\begin{equation}
    \rho \textbf{u}' = \rho \textbf{u}+ \tau
    (\textbf{F}_\sigma+\textbf{F}_w+\textbf{f}),
\end{equation}
where $\textbf{u}'$ is the new velocity used in equation
(\ref{eq}).


We now present the simulation results on a box consisting of $100
\times 50$ lattice units. we fixed the fluid-fluid interaction
 strength $\cal{G}$ = -0.52, fluid-wall interaction strength $w = -0.14$ and
 the relaxation time $\tau = 0.9$. The constant $\rho_0$ in the expression of
  potential $\varphi$ was chosen to be 1.0. At beginning, the fluid was
  homogenous and possessed a density of 0.85. An attractive force
  was assumed to push the fluids to condense into a droplet in the
  middle of the substrate. When the droplet had formed, the
  artificial attractive force was eliminated and a normal droplet
  formed in the middle of the substrate shown as in
  Fig. \ref{droplet}.
\begin{figure}[htb]
  \centering
  {\scalebox{0.35}[0.35]{\includegraphics*[50,220][495,630]
      {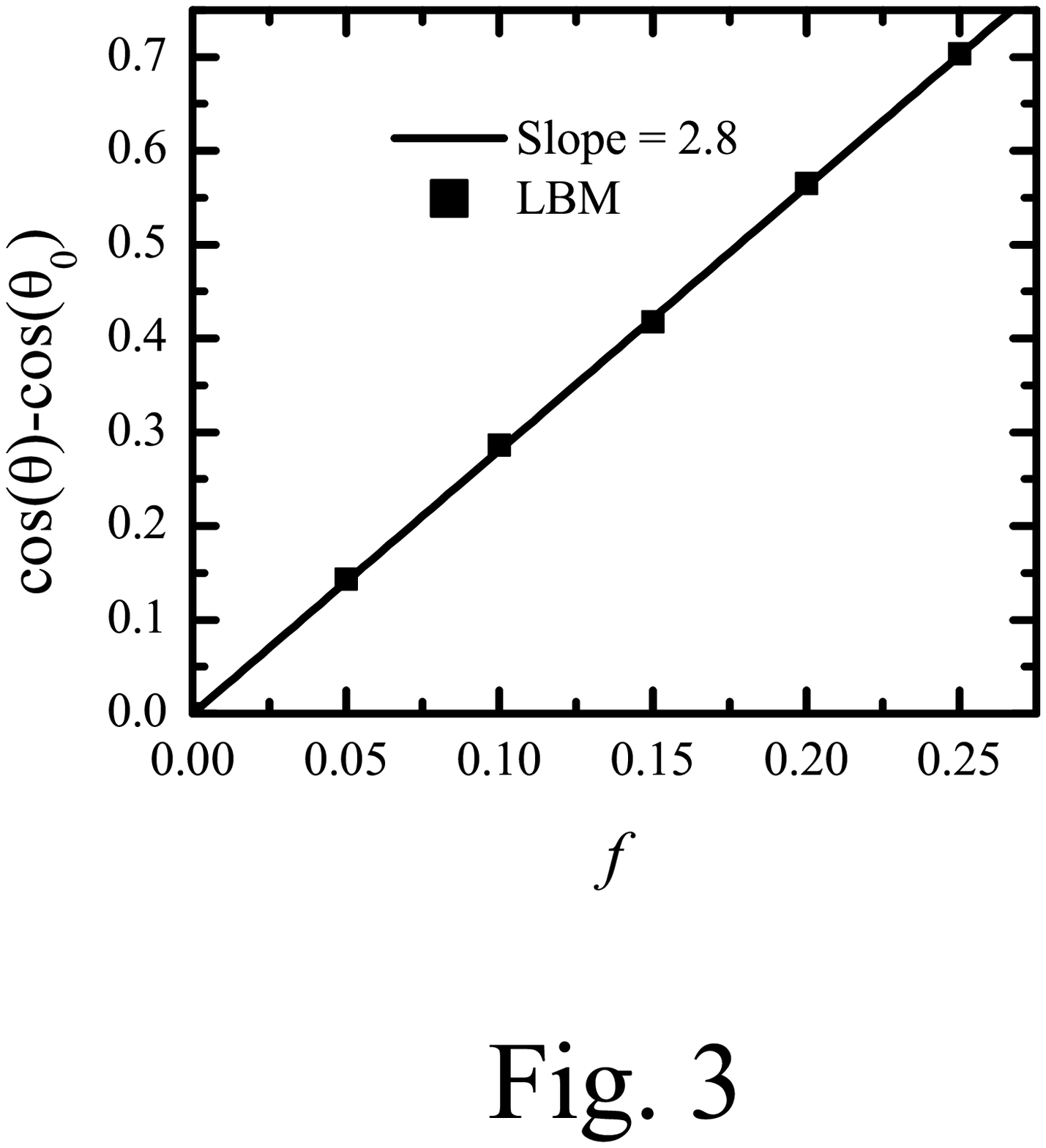}}}
  \caption{$\cos(\theta)-\cos(\theta_0)$ versus electric force $f$. $\theta$
  and $\theta_0$ are the contact angle of the droplet with and without external electric field. }
  \label{angle}
\end{figure}
\begin{figure}[htb]
    {\scalebox{0.4}[0.4]{\includegraphics*[80,325][480,595]
        {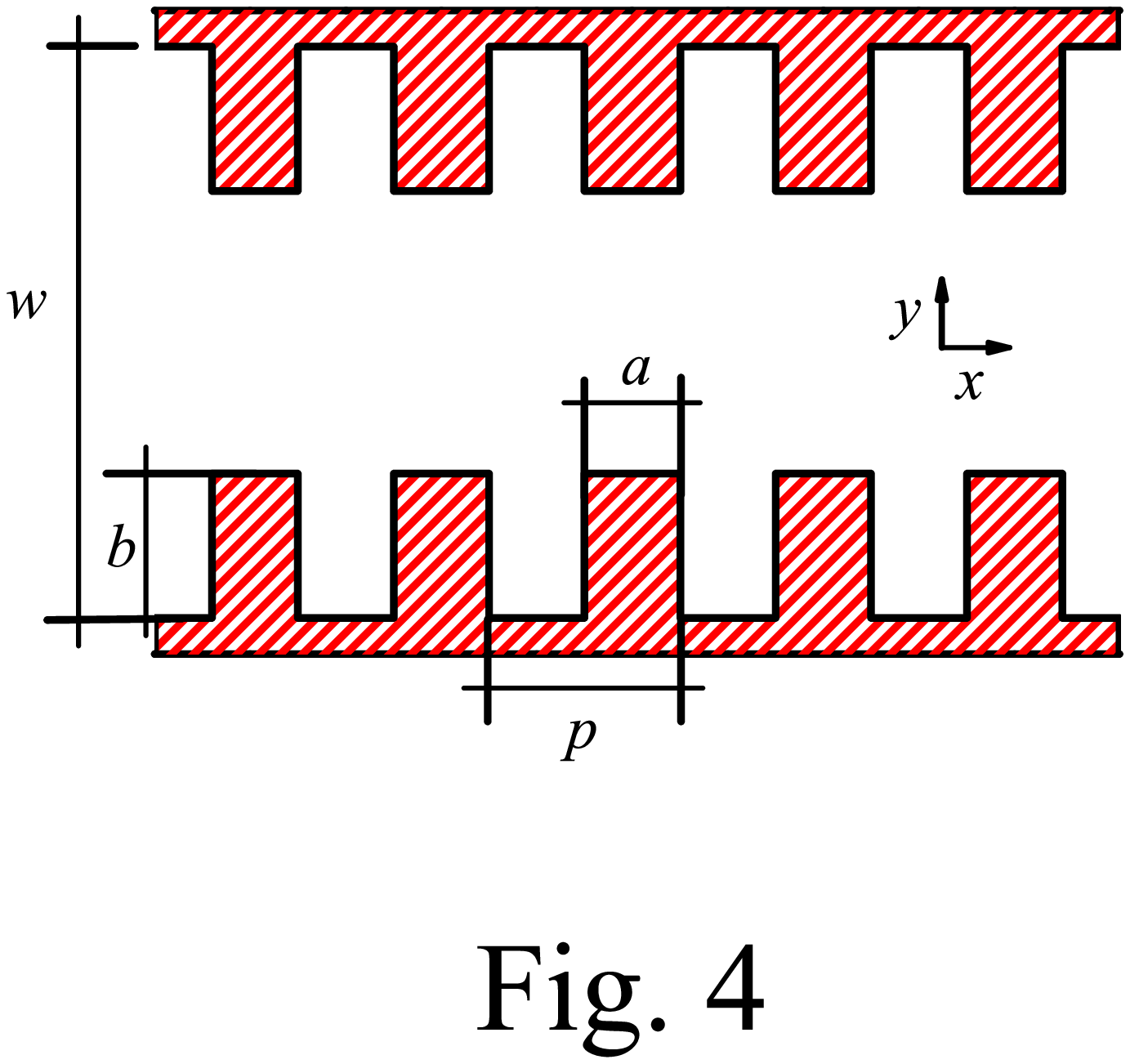}}}
  \caption{Sketch of rough-wall channel.}
  \label{channel}
\end{figure}
\begin{figure}[htb]
  \begin{center}
    {\scalebox{0.28}[0.28]{\includegraphics*[100,280][530,590]
        {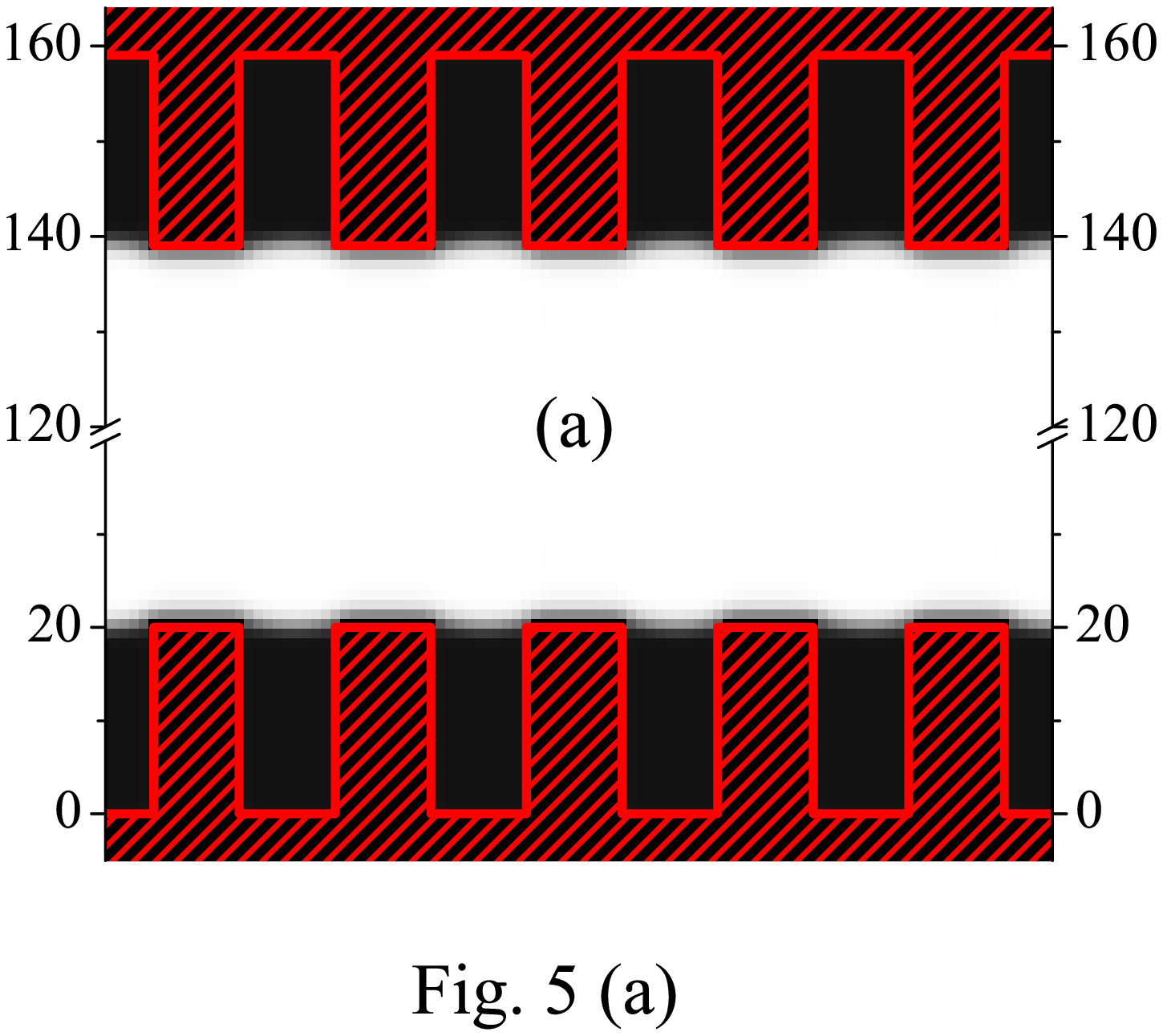}}}
    {\scalebox{0.28}[0.28]{\includegraphics*[100,280][530,590]
        {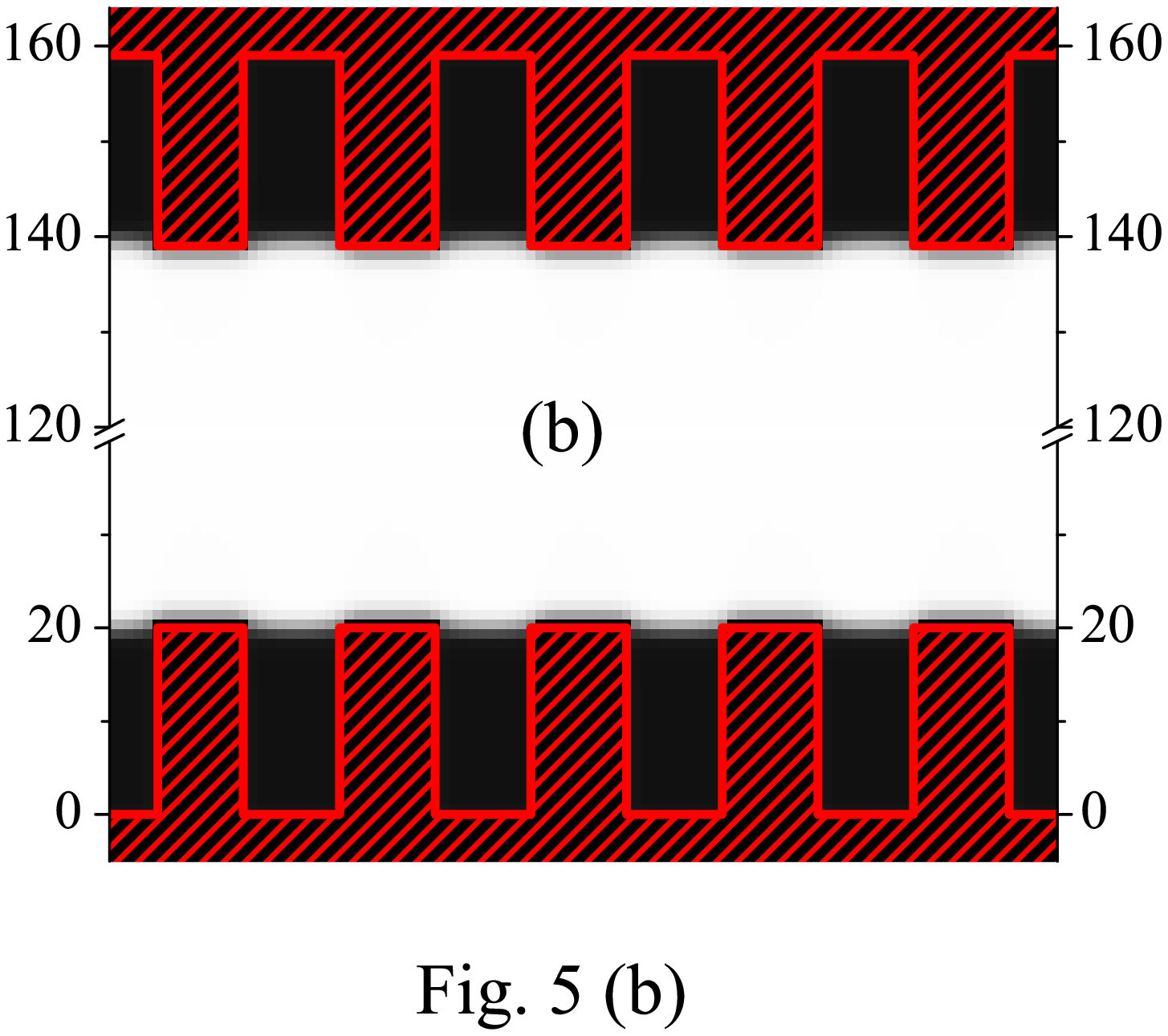}}}
    {\scalebox{0.28}[0.28]{\includegraphics*[100,280][530,590]
        {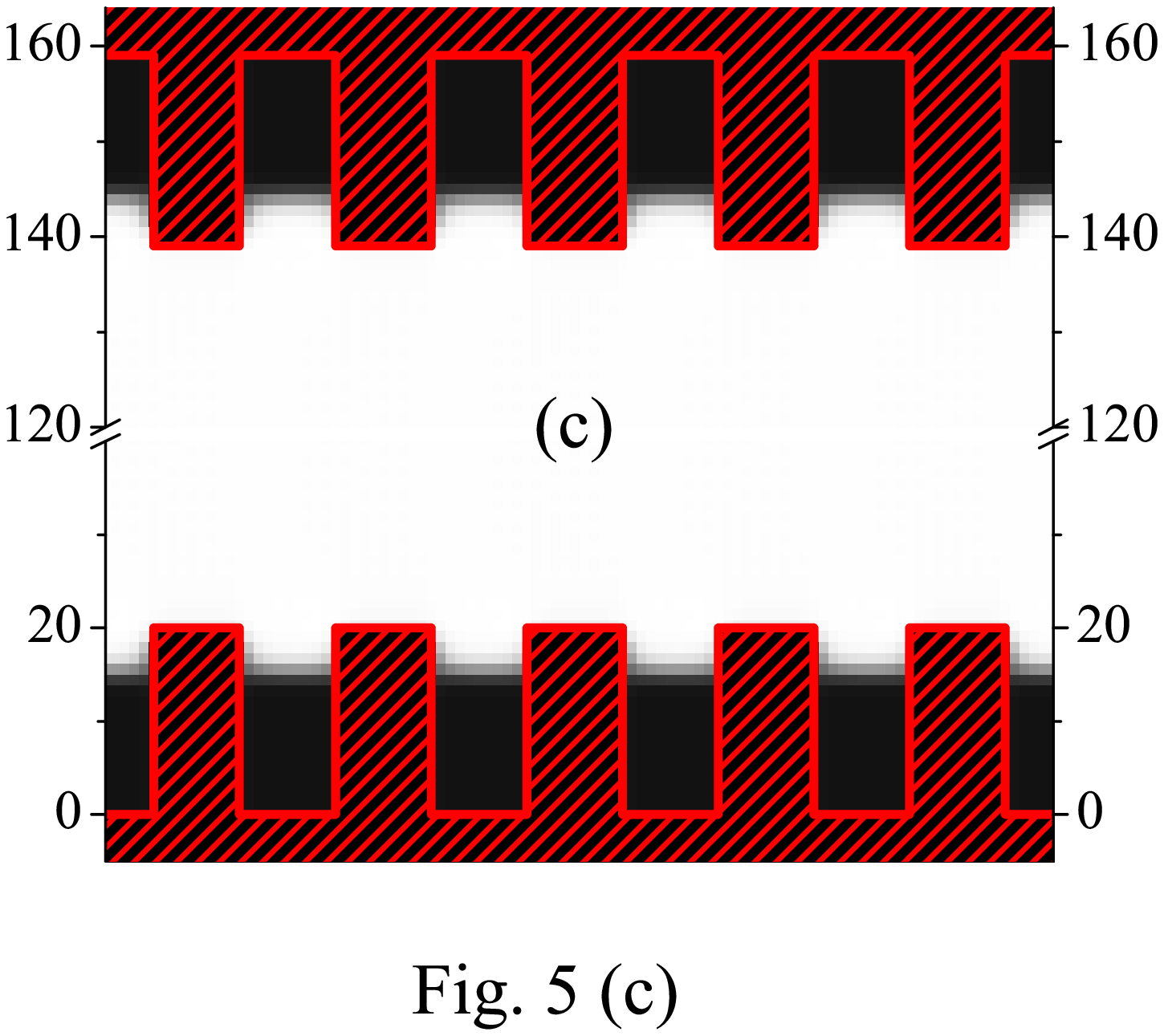}}}
  \end{center}
  \caption{Rough-wall channel with width $w = 160$. The body force pointing
  right is $f_b = 8.0 \times 10^{-6}$. The electric force: (a) $f = 0.0$, (b) $f =0.15$, (c) $f = 0.25$.  }
  \label{channel160}
\end{figure}
Next, an electrostatic force of attraction $f$ was exerted on the
  single layer of fluids near the substrate to simulate
  EWOD. Fig. \ref{angle} shows the contact angle
  changes with the electric force. From this figure we can see
  that $\cos\theta-\cos\theta_0$
  increases proportionally with electric force $f$, consistent with Eq. (\ref{be1}). Thus, this simple model can be used to simulate EWOD.
\begin{figure}[htb]
  \begin{center}
    {\scalebox{0.28}[0.28]{\includegraphics*[100,280][530,590]
        {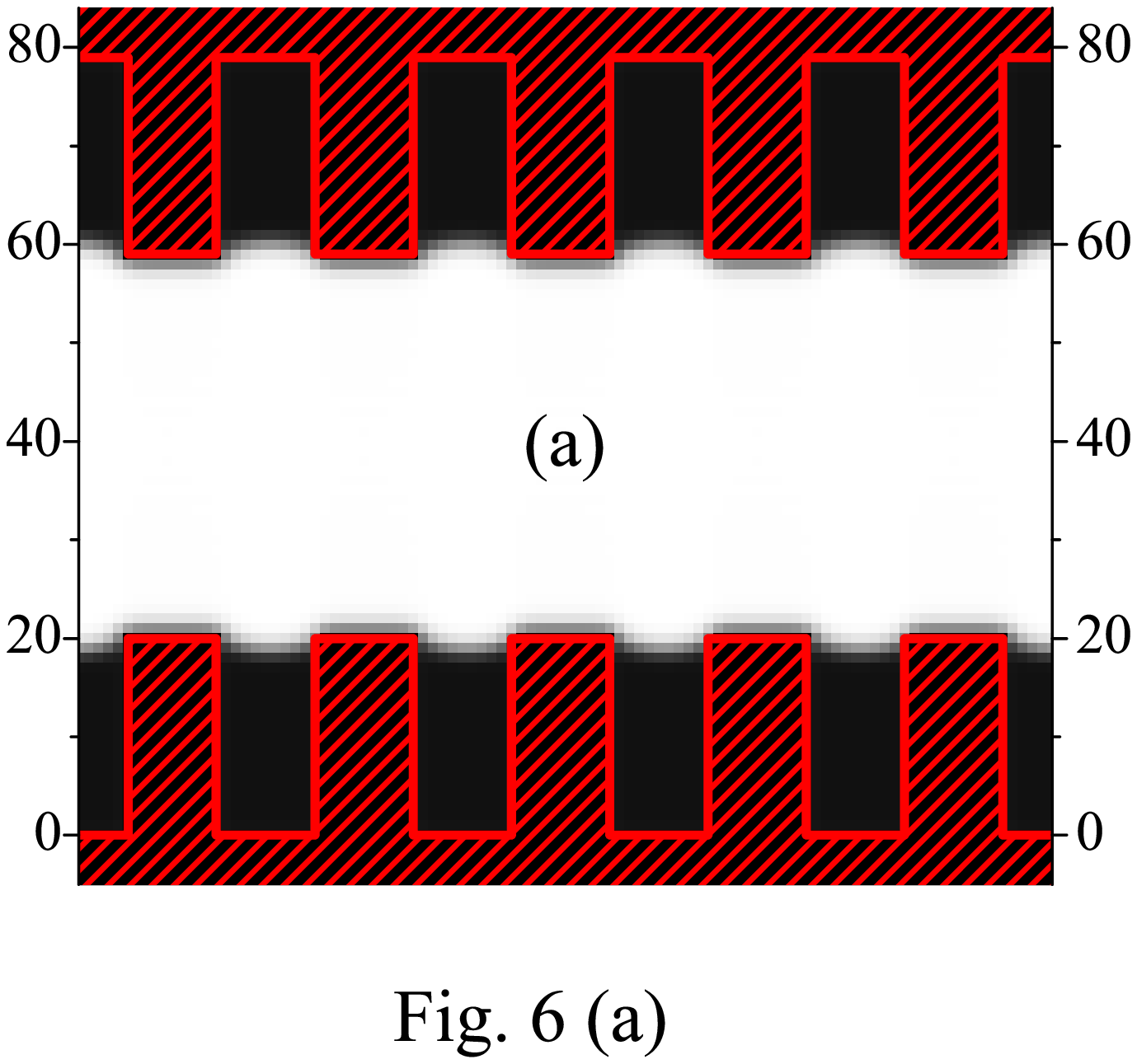}}}
    {\scalebox{0.28}[0.28]{\includegraphics*[100,280][530,590]
        {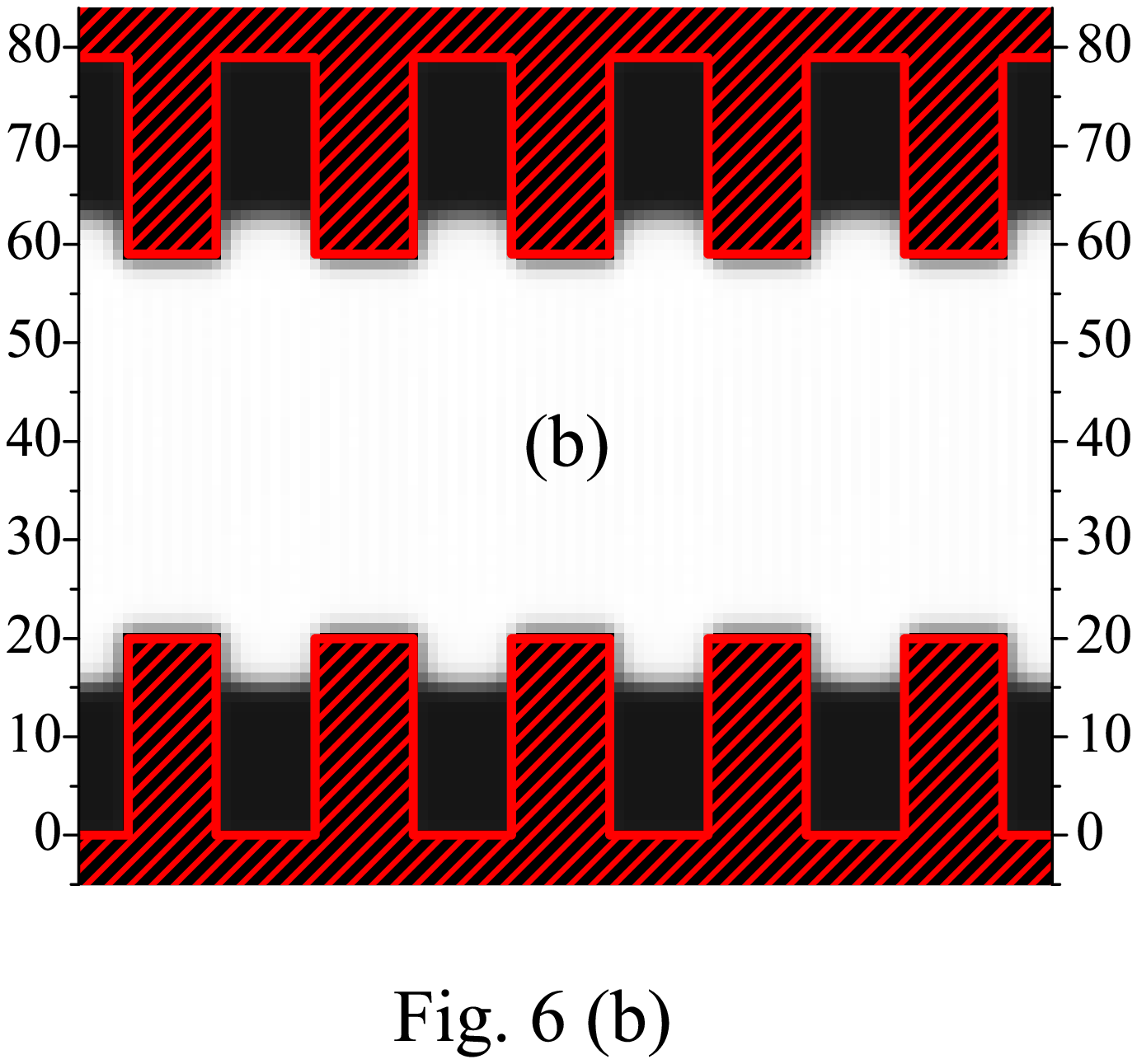}}}
    {\scalebox{0.28}[0.28]{\includegraphics*[100,280][530,590]
        {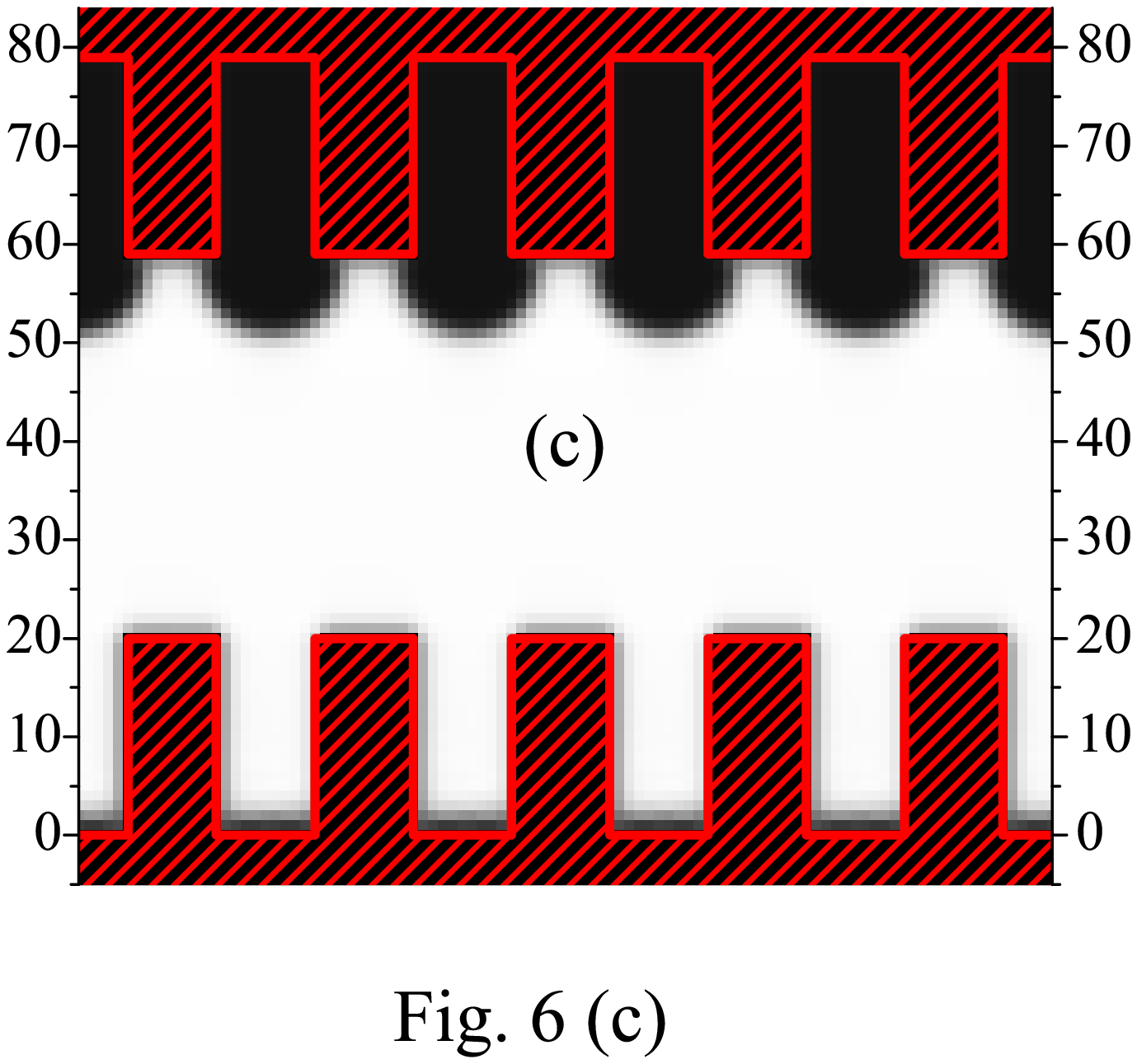}}}
  \end{center}
  \caption{Rough-wall channel with width $w = 80$. The body force pointing
  right is $f_b = 8.0 \times 10^{-6}$. The electric force: (a) $f = 0.0$, (b) $f =0.15$, (c) $f = 0.25$.}
  \label{channel80}
\end{figure}
\begin{figure}[htb]
  \begin{center}
     {\scalebox{0.4}[0.4]{\includegraphics*[40,230][500,590]
        {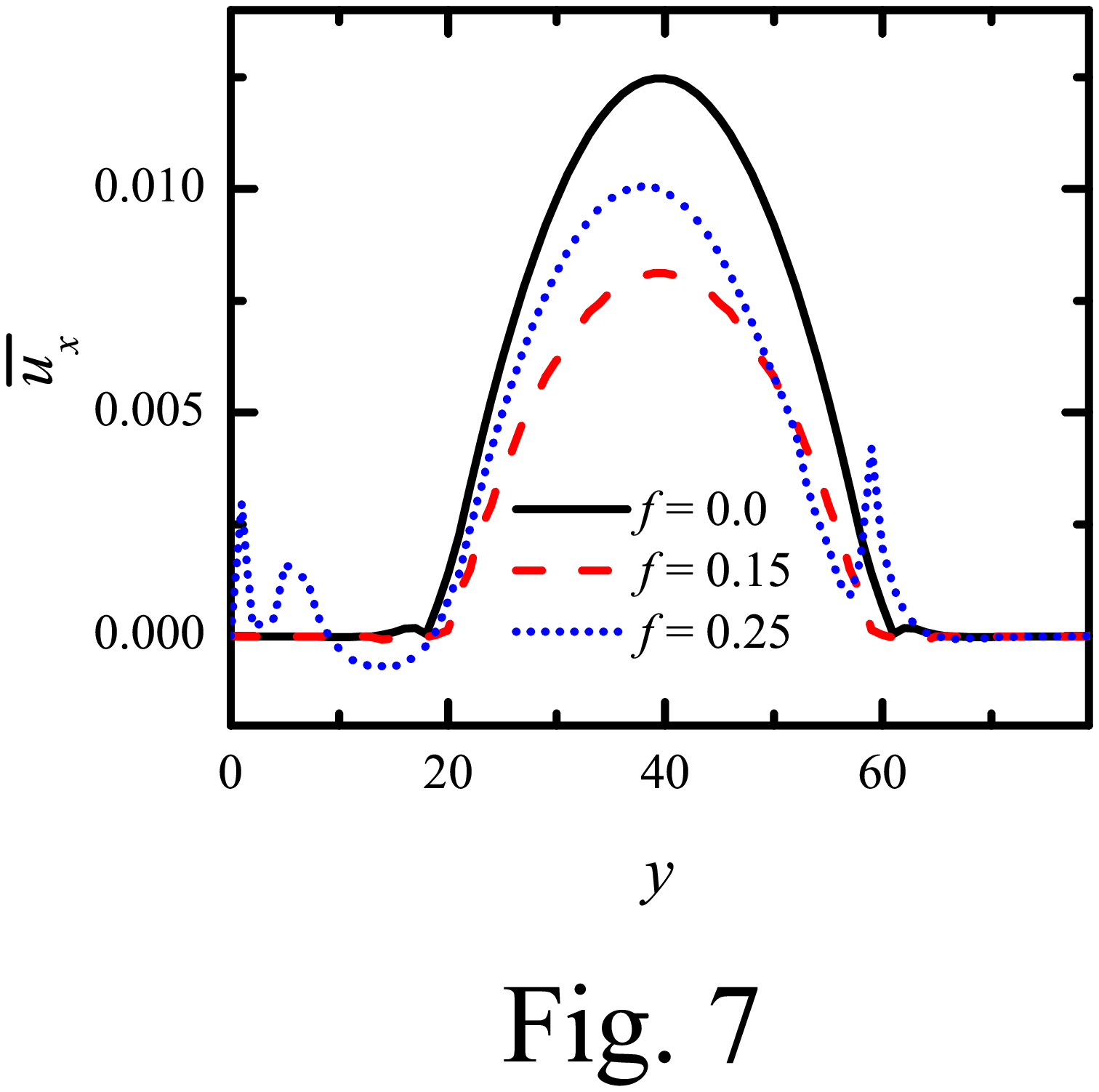}}}
  \end{center}
  \caption{Average of fluid velocity along x-axis  $\overline{u_x}$ in the rough-wall channel with width $w =80$.}
  \label{velocity}
\end{figure}
\begin{figure}[htb]
  \begin{center}
    {\scalebox{0.4}[0.4]{\includegraphics*[60,230][500,600]
        {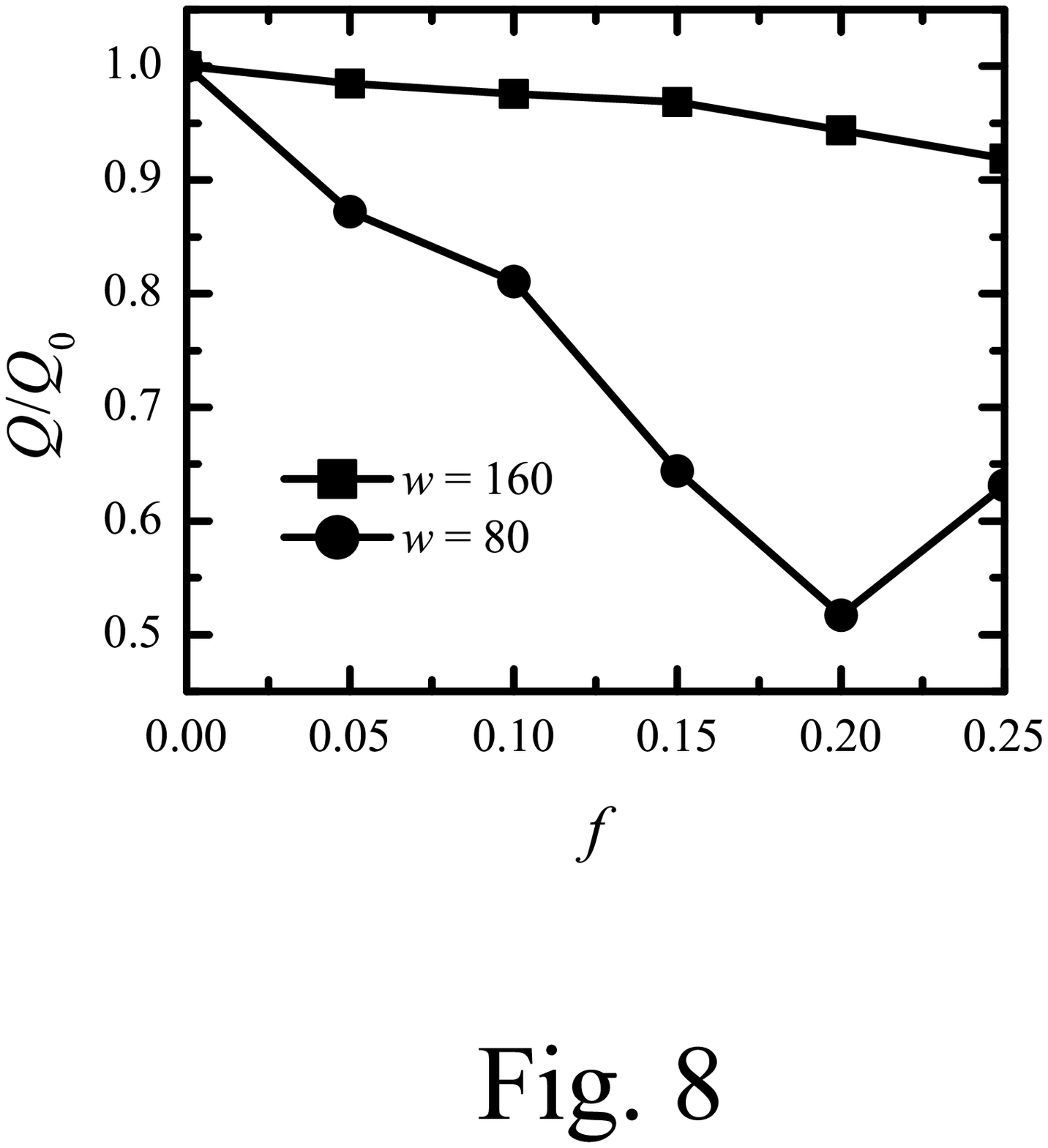}}}
  \end{center}
  \caption{Relative flux of the channel versus electric force.
  $Q$ and $Q_0$ are the electrolyte flux in the channel with and
  without external electric field. The rightward body force exerted
  on the electrolyte is $f_b = 8.0 \times 10^{-6}$.}
  \label{flux}
\end{figure}

For a nanochannel, as the channel size decreases, the
surface-to-volume ratio increases. Therefore, various properties
of the walls, such as surface charges and roughness, greatly
affect the fluid motions in the nanochannels~\cite{Cao06}.
Recently, suffice roughness  effect on micro-flow has been widely
studied from theory \cite{Floryan07}, molecular dynamic (MD) and
other numeric methods \cite{Cao06,Varnik07,Ikeda07} and experiment
\cite{Schultz07}. Here, we focus on the effect of EW on an
electrolyte flow inside a rough-wall channel. Fig. \ref{channel}
shows a sketch of rough-wall channel. The top and the bottom walls
of the channel were both equipped with steps of width $a$ and
height $b$. The centreline distance between two neighbouring steps
was $p$. The channel width was $w$ and the left and right boundary
of the channel were periodic. The electrolyte was insulated from
the channel. Applying a voltage between the channel and the
electrolyte, we can used this system to investigate EW effect on
the electrolyte flow in a rough-wall channel. Fig.
\ref{channel160} shows the electrolyte flowing through the
rough-wall channel under a rightward body force $f_b$ with
different external electric field. From this diagram we can see
that when the applied voltage increases, the electric force $f$
increases ($f \propto U^2$), the electrolyte wets the wall, and
the electrocappirary phenomenon is obviously seen. For a more
narrow rough-wall channel, as shown in Fig. \ref{channel80}, when
the electric force is strong enough, the balance of the system
will be break down and the electrolyte can not stay in the center
of the channel. Fig. \ref{channel80} (c) shows that the
electrolyte are absorbed into one rough-wall of the channel living
only little parts of the electrolyte keep contacting with another
rough-wall. Although the electrolyte keeps flowing rightward, the
contact points keep static, because of small velocity of the flow.

For a narrow channel, not only the properties of the wall affect
the flow very much, to an electrolyte flow, but also the external
electric field or charges. Fig. \ref{velocity} shows the
 average of fluid velocity along x-axis $u_x$ under different external
electric fields. The according dimensionless flux $Q/Q_0$ is shown
in Fig. \ref{flux}. $Q$ is defined as:
\begin{equation}
    Q = \frac{1}{L}\int^{L}_{0}u_x \mathrm{d}x \mathrm{d}y,
\end{equation}
where $L$ is length of the channel and $Q_0$ is the flux without
extenal electric field. From this figure we can see that the
dimensionless flux in a narrow rough-wall channel decreases more
rapidly than in a broad channel. When the electric field is strong
enough, for a narrow rough-wall channel, the electrolyte leans
towards one wall and lives thick vapor layer between the
electrolyte and another wall (see Fig. \ref{channel80} (b)) and
the vapor layer makes the wall slip more (shown in Fig.
\ref{velocity} for $f = 0.25$). Thus the flux increases suddenly
under such external electric field (see Fig. \ref{flux} for $f =
0.25$).

  In conclusion, we introduced a critical density $\rho_c$ to distinguish conductive
   liquid from insulative vapor and an electric force exerted on the conductive liquid
   to improve the LBM to simulate EWOD. The linear property of contact angle confirmed
   the reliability of this moel. Further, we applied this model to
   simulate electrolyte flowing in a rough-wall channel and found
   that the external electric field affect the density
   distribution, velocity and flux greatly for a narrow rough-wall
   channel. For a narrow rough-wall channel, when
   the external electric field is strong enough, the electrolyte
   may be adopted into one rough-wall and departure another
   rough-wall, reducing the whole resistance and resulting a
   suddenly increasing of flux in the channel.

   This work was supported by grants from Chinese Academy of
Sciences, the National Science Foundation of China under grants
Nos. 10474109, 10447001 and 10674146, the National Basic Research
Program of China under grant Nos. 2006CB933000 and 2006CB708612
and Shanghai Supercomputer Center of China. The Guangxi Science
Foundation Nos. 0640064.

\end{document}